\documentclass[twocolumn]{aastex63}

\usepackage{amsmath}

\usepackage{hyperref}
\hypersetup{colorlinks=true,
linkcolor=red,
citecolor=blue,
filecolor=cyan,
urlcolor=magenta,}

\renewcommand{\deg}{^\circ}

\newcommand{\eg}{e.g.\ }

\def\up#1{{$^{#1}$}}
\newcommand{\linclacc}{$L_{inc}/L_{acc}$}
\newcommand{\gimtd}{{\it GIM2D}}
\newcommand{\imfit}{{\it imfit}}
\usepackage{array}
\newcolumntype{L}[1]{>{\raggedright\let\newline\\\arraybackslash\hspace{0pt}}m{#1}}
\newcolumntype{C}[1]{>{\centering\let\newline\\\arraybackslash\hspace{0pt}}m{#1}}
\newcolumntype{R}[1]{>{\raggedleft\let\newline\\\arraybackslash\hspace{0pt}}m{#1}}

\accepted{January 20, 2024}

\submitjournal{AJ}

\shorttitle{SN\,Ia Distribution in Host Galaxies}
\shortauthors{Pritchet et al.}

\begin{document}

\title{The Spatial Distribution of Type Ia Supernovae within Host Galaxies}

\correspondingauthor{Chris Pritchet}
\email{pritchet@uvic.ca}

\author{Christopher Pritchet}\
\author{Karun Thanjavur}
\affiliation{Dept. of Physics and Astronomy, University of Victoria, 3800 Finnerty Road, Victoria, BC, Canada}

\author{Connor Bottrell}
\affiliation{International Centre for Radio Astronomy Research, University of Western Australia, 35 Stirling Hwy, Crawley, WA 6009, Australia}

\author{Yan Gao}
\affiliation{Dept. of Physics and Astronomy, University of Victoria, 3800 Finnerty Road, Victoria, BC, Canada}

\begin{abstract}

We study how type Ia supernovae (SNe\,Ia) are spatially distributed within their host galaxies, using data taken from the Sloan Digital Sky Survey (SDSS). This paper specifically tests the  hypothesis that the SNe\,Ia rate traces the $r$-band light of the morphological component to which supernovae belong.  A sample of supernovae is taken from the SDSS SN Survey, and host galaxies are identified. Each host galaxy is decomposed into a bulge and disk, and the distribution of supernovae is compared to the distribution of disk and bulge light. Our methodology is relatively unaffected by seeing. We find that in disk light dominated galaxies, SNe\,Ia trace light closely. The situation is less clear for bulges and ellipticals because of resolution effects, but the available evidence is also consistent with the hypothesis that bulge/elliptical SNe\,Ia follow light.  

\end{abstract}


\section{Introduction} 
\label{sec:intro}

Type Ia supernovae  (SNe\,Ia) are the thermonuclear detonations of carbon-oxygen white dwarfs (CO WDs) \citep{Hoyle1960,Arne69, Nomo82a}, or possibly oxygen-neon white dwarfs \citep{Marq2015,Augu2019,GalY2022}, which, by accretion of material from a binary companion, reach a critical mass and explode. 
\citet{Livi18} and others 
broadly classify the pathways to explosion into two channels:  {\it Single Degenerate}  \citep{Whel73}, and  {\it Double Degenerate}  \citep{Iben84}. In the former case, a WD accretes material from a non-degenerate, close binary companion through Roche-Lobe transfer or through winds from a companion, until the WD reaches some critical mass and explodes. In the latter model, a binary WD system reaches a critical mass via disruption and accretion of one companion, or by coalescence. Double degenerate SN\,Ia ignition may even occur as a result of head-on collisions in dense stellar environments. The critical mass for explosion is usually taken to be the maximum mass of a WD, the \citet{Chandrasekhar1931} mass, $M_{Ch} \simeq 1.4$ M$_\odot$; however there exist mechanisms by which  sub-Chandrasekhar mass explosions can occur (see \citealt{Livi18, Ruit20, Maoz14} for these and other progenitor scenarios). 

Each progenitor mechanism has strengths and weaknesses \citep{Livi18}, and at the present time it is unclear which of the many progenitor scenarios is dominant. (To quote Livio and Mazzali, ``{\it all} the existing progenitor scenarios encounter difficulties".) Nor do there exist any observations of pre-explosion SNe\,Ia progenitors, or for that matter of post-explosion stellar remnants of SNe\,Ia (which might be expected for the single degenerate channel). This is the SN\,Ia ``progenitor problem''.

Given the observational challenges of identifying individual SNe\,Ia progenitor systems, one alternate approach to the progenitor problem is to study how SNe\,Ia are correlated with the environment in which they form or are found. For example, one can compare the radial distribution of supernovae with the radial distribution of stars. Most radial distribution studies have, however, been concerned with core collapse (CC) SNe, which are known to trace the spiral structure of their hosts (e.g. \citealt{Maza76}). More recent studies have consistently shown that CC SN rates are closely related to near ultra-violet (nUV) and H$\alpha$ luminosities of hosts, both of which are indicative of active formation of massive stars (e.g. \citealt{Fruc06,Jame06,Kell08,Rask08,Habe10,Pere14,Hako16,Aram16,Hako17,Chak18,Audcent-Ross2020,Schu21}).

In contrast, little is known about the detailed spatial distribution of SNe\,Ia within their hosts. It is to be expected that SNe\,Ia will broadly track mass and light density; however, a complicating factor is that SNe\,Ia have a wide range of delay times (where the delay time is the time between the onset of the main sequence phase of a SN\,Ia progenitor, and its actual SN explosion). SNe\,Ia from the most massive progenitors may be found within a kpc of their formation region, but this is not true for most SNe\,Ia.

Compared to CC SNe, SNe\,Ia show a much broader spatial distribution relative to the spiral arms \citep{Aram16}, and are also further from the plane of the disk \citep{Hako09, Hako12, Hako14}. Taking the spiral arms and the disk to be the locations of active star formation, the broad conclusion from this series of papers is that SNe\,Ia occur at significant distances from the locations of star formation, as expected for long delay times. 

Similarly, \citet{Ande15} have applied a method of pixel statistics developed in \citet{Ande08} and \citet{Fruc06} to show that SN\,Ia distributions do not correlate with the nUV and H$\alpha$ light associated with active star formation, thus arguing against a dominant young or `prompt' progenitor pathway. They found the best correlation with the underlying B-band light, indicative of an intermediate age progenitor population. On the other hand, in early type host galaxies, studies have shown that SNe\,Ia track the light distribution of their old, evolved stellar populations, indicating long delay times, of order several gigayears \citep{Fors08,Bark19,Hako20,Audc20}. 

How would the SNe\,Ia rate be expected to depend on host galaxy surface brightness? To answer this question, consider the delay-time distribution, DTD(t), which is defined as the number of supernovae from a burst of star formation as a function of the age of the burst (normalized per unit mass of the burst). There exist arguments that $DTD \sim t^{-1}$, where $t$ is the age of a stellar population  \citep{Maoz14,Heringer2019}. In this case, and somewhat remarkably, the SNe\,Ia rate per unit $r$-band luminosity, $SNR/L_r$, can be shown to be nearly constant over a wide range of stellar population color, metallicity, or star formation history \citep{Heringer2016,Heringer2019}. The reason for this is the (somewhat fortuitous) cancellation of the effects of M/L (which increases with increasing color) and SNR (which decreases with increasing age and hence color). However, for a steeper DTD$\sim t^{-1.5}$ there is some weak $g-r$ host color dependence of $SNR/L_r$. Given that the DTD power-law index probably lies between $-1$ and $-1.5$ \citep{Heringer2019}, and that the color gradients in bulges and disks are small, the {intrinsic} SN\,Ia rate and light intensity should track each other quite closely, at least within a single morphological component. 

The goal of our paper is to test the hypothesis that SN\,Ia track light, using a { subsample} of the $\sim$2000 SNe\,Ia from the SDSS-II Stripe 82 Supernova Survey \citep{Sako18}. This well-characterized supernova survey includes host galaxies with a wide variety of morphological types and star formation histories, making them ideal for testing whether SNe trace light. Typical SN\,Ia  host galaxies in this survey have redshifts $z \simeq 0.2-0.3$, apparent magnitudes $r \simeq 18-20$, and sizes of a few arcsec. Seeing and optics blur the distribution of light in such hosts, so it is necessary to fit the underlying distribution of light with one or two component models, prior to comparing with the distribution of supernovae (whose positions are less affected by image quality considerations -- e.g. Appendix \ref{subsec:xyerr}).

{However, it is necessary to differentiate between the {\it true, intrinsic} SN\,Ia rate, and what is observed. The observed rate in a particular galaxy depends on a variety of astrophysical effects (stellar age being crucial; metallicity and binary frequency also play a role), galaxy orientation (which modulates the effects of dust obscuration), and observational incompleteness. While it is difficult to estimate the astrophysical and geometrical corrections to derive an intrinsic SN\,Ia rate, it is important to  note  that our analysis is sensitive only to {\it radial gradients} in the $SNR/L$ ratio. In addition, $SNR/L$ is analyzed in different components (bulge and disk) separately. Both of these analysis techniques will minimize (though not eliminate) astrophysical and orientation effects. Observational incompleteness is amenable to numerical experiments, and these are discussed both in the analysis sections (\S\S\ref{sec:disks}--\ref{sec:bulge+disk}) and also Appendix \ref{subsec:completeness}. }

We have organized the paper as follows: In Section \ref{Cats}, we describe the SDSS-II stripe 82 Supernova Survey \citep{Sako18} used for this study, and the quantitative metric used to identify SN\,Ia host galaxies. \S\ref{subsec:hostfitting} describes  two independent galaxy morphology fitting methods, \gimtd~ \citep{Sima02} and \imfit~ \citep{Erwi15} that we used to map the underlying host galaxy light distributions. We test for a correlation between the SN\,Ia distribution and the  light distribution of the host galaxies using the \linclacc~ metric, which is defined and described in \S\ref{subsec:linclacc}. Our large sample size permits us to study the \linclacc~ distribution in statistically significant subsets of our primary sample, as described in \S\ref{subsec:sample}. An initial analysis is presented in \S\ref{sec:firstlook}. We provide results for disk dominated hosts in \S\ref{sec:disks}, for bulge dominated hosts in \S\ref{sec:bulges}, and for the mixed bulge+disk sample in \S\ref{sec:bulge+disk}. \S\ref{sec:discussion} is devoted to a discussion of our main findings, and our conclusions are summarized in \S\ref{sec:conclusions}. {The Appendices  provide more details on the fitting of galaxy structural parameters (and their associated errors), on the calculation of \linclacc, and on biases in the light profile fitting (especially incompleteness effects).}

\section{Type Ia Supernovae and Host Galaxies}
\label{Cats}

\subsection{SN Ia Sample}

Our SN\,Ia sample is taken from the Sloan Digital Sky Survey (SDSS) Stripe 82 Supernova Survey
\citep{Sako14,Sako18}, henceforth referred to as the \textit{S82-SNe} survey. The survey strategy was to use SDSS to repeatedly observe the same region of sky every alternate night during a three month observing window (Sep-Nov) over three years (2005 to 2007). The imaging was conducted within Stripe 82, covering an area of $\sim$275 deg$^2$ in a $2.5 \deg$ wide belt centered on the celestial equator and spanning $-50 \deg \le$ RA $\le 59 \deg$. The survey efficiency and the selection biases have been investigated in detail by \citet{Dild08, Dild10a}.

A total of 10258 transients were discovered during the full survey. The light curves of the transients were classified with the \textit{Photometric SN IDentification} (PSNID) \citep{Sako11}. 
For our analysis, we use only the three broad groups of Type Ia supernovae: SNIa (499 spectroscopically confirmed Type Ia SNe); zSNIa (824 photometrically classified SNe\,Ia with spectroscopic redshifts from their host galaxies); and pSNIa (624 photometrically classified SNe\,Ia with photometric redshifts only). The total number of SNe\,Ia is 1947. We omit a fourth class of uncertain identifications (SNIa?, 41 objects) from our study. {The completeness of the SN\,Ia sample is discussed in \S\ref{subsec:sample} and Appendix \ref{subsec:completeness}.}

\subsection {Supernova - Host Galaxy Matching}
\label{subsec:hostmatching}

In the crowded galaxy field typical of a deep imaging survey, the galaxy with the closest angular distance (CAD) to a supernova is not always its host. \citet{Gupt16} and \citet{Gagl21} discuss the challenges of identifying the host galaxies of transients in large imaging surveys such as the \textit{S82-SNe}. 

We have adopted an algorithm similar to those of \citet{Sull06} and \citet{Sako18} for SN--host galaxy matching. Photometric and geometric properties of galaxies are taken from the \textit{SDSS-DR7} photometric catalogs \citep{Abaz09} for the Stripe 82 footprint. We define a dimensionless quantity $\rho_{25}$, which is the projected separation of a supernova from a galaxy, measured in units of the $r$-band $25$ mag arcsec$^{-2}$ isophotal radius at the position angle of the supernova ($d/d_{DLR}$ in the terminology of \citet{Sako18}). The host galaxy of a particular supernova is taken to be that galaxy with a minimum value of $\rho_{25}$.

We select only hosts whose $\rho_{25}$ separations lie below $\rho_{25}^{crit} =1.959$, for which 8\% of randomly positioned (unhosted) SNe would be assigned to a host. A simulation shows that $\sim 0.2\%$ of objects would be assigned to the wrong host. Most importantly, 97\% of our host identifications agree with those of \citet{Sako18} for $r_{host}<20$ mag. Also of interest is the fact that 86.5\% of our host identifications for the full host catalog agree with the galaxy with the closest angular distance. 

\section{Analysis}
\label{sec:anal}

Our technique for comparing SNe\,Ia and host galaxy light (the \linclacc\  method described in \S \ref{subsec:linclacc}) relies on obtaining an estimate of the intrinsic light distribution of host galaxies, affected as little as possible by seeing. To do this, we fit a seeing-convolved model to each host galaxy, as described below.

\subsection {Host Fitting}
\label{subsec:hostfitting}

SDSS SNe\,Ia hosts at $z \simeq 0.2$ are fairly similar (in both median luminosity, and also, as we shall see, median bulge to total light ratio) to the nearby spiral galaxy M31. Scaling from the properties of M31 \citep{Courteau2011}, we find that M31-like galaxies at $z=0.2$ have a bulge half-light radius (HLR) of 0.3 arcsec, and a disk HLR of 3 arcsec. This illustrates the importance of seeing effects on host galaxies (and especially their bulge components) at the typical redshifts of SDSS SNe. To test whether SNe trace galaxy light, it is necessary to compare the spatial distribution of supernovae with the {\it true, intrinsic} light distribution of galaxies, unaffected by seeing\footnote{While seeing blurs the core light of SN hosts, it affects the distribution of SNe less, because the {\it centroids} of point sources such as SNe can be measured to $\pm 0.1$ arcsec or better, much smaller than the seeing disk. This assumption is discussed further in Appendix \ref{subsec:xyerr}.}.

Each host galaxy was fitted using two different galaxy light modelling programs, to provide added confidence in the fits. The programs used were \gimtd\ \citep{Sima02}, and \imfit\ \citep{Erwi15}. These programs fit multiple component galaxy models to seeing-degraded image data, convolving each model with a pre-determined seeing profile to match the observations. \gimtd\ uses the Metropolis algorithm \citep{Metr53} to search for the best fit in parameter space; \imfit\ allows the use of a variety of best-fit search algorithms, including a genetic algorithm. 

The fitting functions used with each program were one or two elliptical S\'ersic \citep{Sers63} profiles of the form

\begin{equation}
ln (I/I_e) = b_n [(r/r_e)^{1/n}-1],
\label{eq:Sersic}
\end{equation}

\noindent
with profile parameters $r_e$ (half-light radius), $I_e$ (surface brightness at $r=r_e$), S\'ersic index $n$, and with constant (i.e. spatially invariant) shape and orientation parameters (ellipticity $b/a$ and position angle $PA$). The value of $b_n$ is calculated as in \citet{Grah05}. 

Most of the fits were made with a two component (disk plus bulge) model, where the disk is exponential ($n=1$), the bulge has a de Vaucouleurs profile ($n=4$), and $r_e$, $I_e$, $PA$, and $b/a$ are fitted separately for each component\footnote{For an exponential disk ($n=1$), half light radius $r_e$ and exponential disk scale-length $\beta$ are related through $r_e=1.678 \beta$.}. At the typical redshift of SDSS SNe\,Ia ($z \simeq 0.2-0.3$), a typical bulge is only a few pixels in effective radius, and hence it is not feasible to fit for the bulge S\'ersic index $n_{bulge}$ for the two component models \citep[see][\S4.5]{Bott19}. The assumption $n_{bulge}=4$ is discussed further in Appendix \ref{subsec:bulgen}. In addition to the two-component fitting, we made an additional fit to each host using a single S\'ersic component, allowing its index $n$ to vary.

We used $r$-band Stripe 82 stacked images  \citep{Anni14} for both \gimtd\ and \imfit\ fits. The PSFs and other properties of these stacked images are discussed in Appendix \ref{subsec:PSF}. More information on the use of \gimtd\  and \imfit\ is provided in Appendices \ref{subsec:gim2d} and \ref{subsec:imfit}, {where the determination of errors in the fitted model parameters is also discussed. The most important error parameter is $\sigma_{B/T}$, which is used in the definition of the primary sample (\S\ref{subsec:sample}). The reliability of the models is tested by comparing the \linclacc\ results of the two programs in Appendix \ref{subsec:comparisonfit}.} 

Biases in the profile fitting are discussed in Appendix \ref{sec:biases}. Most of the results that we discuss in \S\S \ref{sec:disks}--\ref{sec:bulge+disk} refer to \gimtd; the results with \imfit\ are similar, except where noted.

\subsection {Do Supernovae Trace Light? The \linclacc\  Metric}
\label{subsec:linclacc}

It is a remarkable fact that, for a stellar population of a given color, the SN\,Ia rate per unit $r$-band luminosity, $SNR/L_r$, is almost independent of age and star formation history (\citealt{Heringer2016,Heringer2019}; see \S\ref{sec:intro}). One would therefore expect the distribution of supernovae to follow light quite closely. 

\begin{figure}
\epsscale{1.4}
\plotone{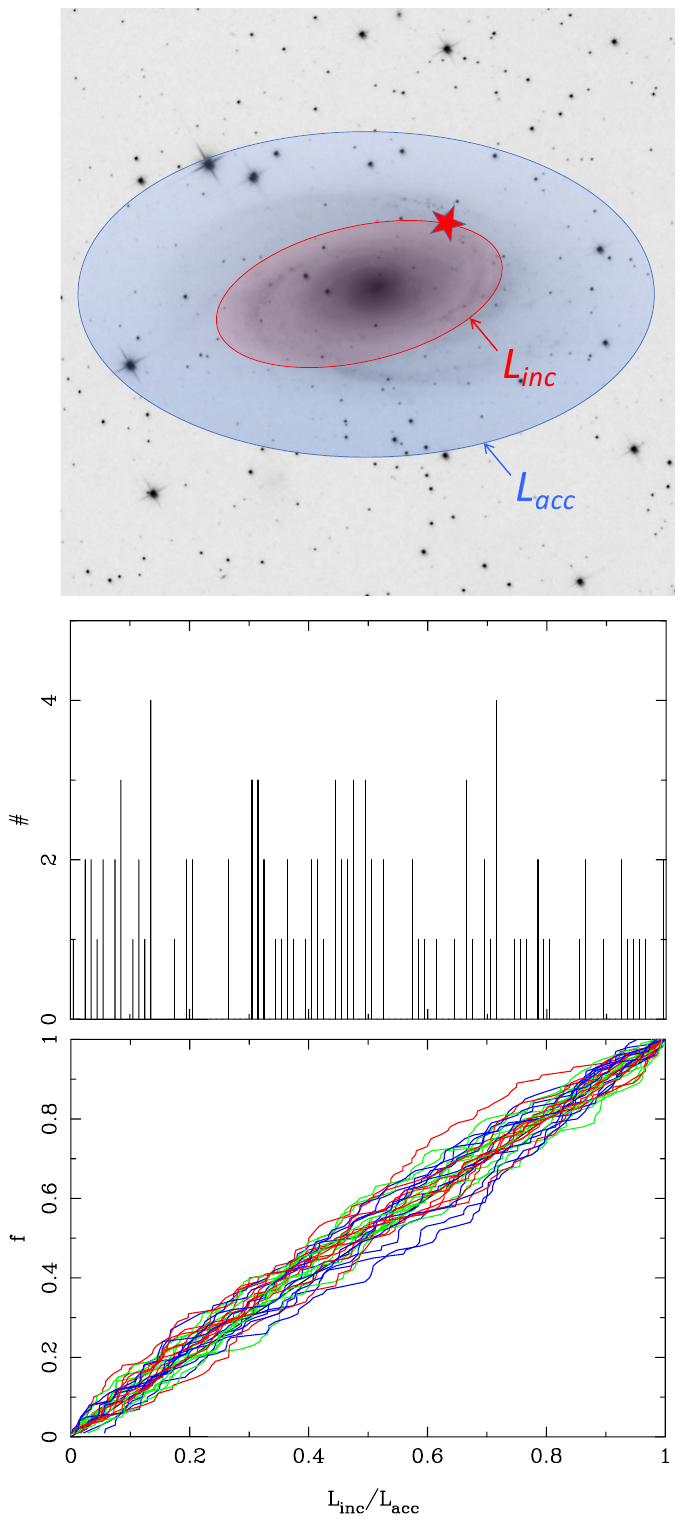}
\caption {Definition of $L_{inc}$ and $L_{acc}$. The {\it upper panel} shows a SN (in red) superimposed on its host galaxy. The red shaded area is the area used to calculate $L_{inc}$; its boundary is defined by an isophote passing through the SN; this isophote is not necessarily elliptical. $L_{acc}$ is computed within an elliptical boundary with semi-major axis $a=1.959 a_{25}$ and semi-minor axis $b=1.959 b_{25}$. This is the area within which a SN can be matched to its host; see \S \ref{subsec:hostmatching}. Note that $L_{inc}$ and $L_{acc}$ are computed using the fitted model light distribution. The {\it middle panel} shows one simulation of 100 SNe in which SNe trace light (bin size 0.01). The {\it lower panel} shows 30 simulations of 100 SNe, plotted as cumulative distributions. }
\label{fig:linclacchowto}
\end{figure}

To test this hypothesis, we use the \linclacc\ technique (\eg \citealt{Fruchter2006, Raskin2008, Audcent-Ross2020}, and references therein). We take the host galaxy isophote passing through the coordinates of a SN, and sum up the fitted light, $L_{inc}$, included (enclosed) within that isophote (see Fig. \ref{fig:linclacchowto}, upper panel).  We also sum the ``accessible" fitted light, $L_{acc}$, included within the elliptical aperture that was used for SN--host matching (with axes $1.959 a_{25}$ and $1.959 b_{25}$ - see \S\ref{subsec:hostmatching}). ($L_{acc}$ is typically 80-90\% of the total light of the galaxy.) \emph{Since $L_{inc}/L_{acc}$ is a ratio of enclosed light}, objects that are stochastically sampled from the light distribution are uniformly distributed in $L_{inc}/L_{acc}$, {\it if} supernovae follow light. It follows that multiple supernovae in {\it  different} host galaxies will also be uniformly distributed as in Fig. \ref{fig:linclacchowto}, and the cumulative distribution of supernovae with \linclacc\  is linear (Fig. \ref{fig:linclacchowto}, bottom panel). Simple statistical tests  (\S\ref{subsec:stattests}) can be applied to Fig. \ref{fig:linclacchowto} to test the null hypothesis that supernovae trace light.

$L_{inc}$ and $L_{acc}$ are calculated from the fitted \gimtd\ and  \imfit\ models. The calculations are complex because the $L_{acc}$ aperture does not necessarily have the same shape and orientation as the host galaxy components, and because the $L_{inc}$ isophote is not elliptical for a two component model (unless both components have exactly the same $b/a$ and $PA$). For this reason we used two independent programs (written by CP and KT) to calculate and check the results; these programs agree to better than $\pm 0.001$ in \linclacc.

Although \linclacc\  was described above in terms of total light, one can use it to test whether supernovae follow the light of only one component, by computing \linclacc\  using bulge light or disk light alone. For the total light calculation one can also allow for a different $SNR/L_r$ in the bulge and disk.

Appendix \ref{subsec:approxlinclacc} discusses two shortcuts that can be used in the calculation of \linclacc; Appendix \ref{subsec:comparisonfit} compares the \linclacc\  values computed using \gimtd\ and \imfit.

\subsection {Statistical tests}
\label {subsec:stattests}

The null hypothesis that SNe follow light can be tested with the well-known Kolmogorov-Smirnov test applied to the cumulative \linclacc\  diagram (Fig. \ref{fig:linclacchowto}). While this is a simple calculation, it is mainly sensitive to deviations in the middle of the range $0 < L_{inc}/L_{acc} < 1$. We therefore also use the Anderson-Darling test, which has greater sensitivity overall, but especially at the ends of the distribution. (The reader is referred to \citealt{Feigelson2020} for a discussion and comparison of these two tests.) We have tested our probability calculations with Monte Carlo simulations, and find them to be in excellent agreement with statistical tables for these 2 tests. {We have also experimented with bootstrapping the distributions of $L_{inc}/L_{acc}$ values, and find that the bootstrapped KS test results are in good agreement with the original KS test  provided that the data supports $SNR \propto L$. }

\subsection {Primary Sample}
\label {subsec:sample}

We now consider the properties of the host galaxies and supernovae, in order to determine which objects to use in our analysis. 

\begin{figure}
\plotone{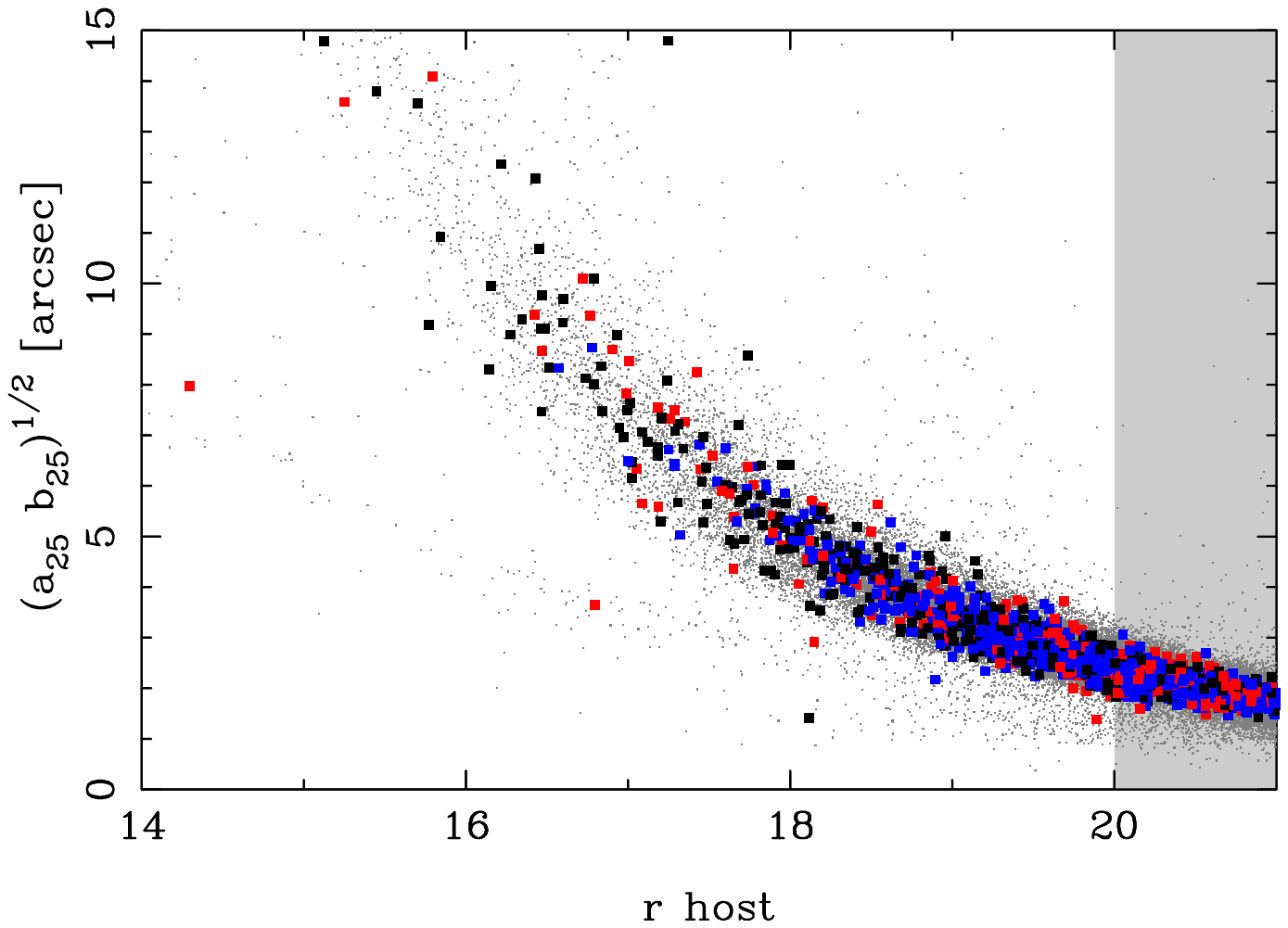}
\caption {Galaxy isophotal size in arcsec  vs. host $r$ band magnitude. The size is the circularized radius of the $\mu_r=25$ mag/arcsec\up{-2} isophote as given in SDSS DR7. The small {\it grey points} are a sample of 93000 galaxies brighter than $r=21$ mag in Stripe 82 with $0^o<\alpha<10^o$. {\it Squares} are host galaxies, colour coded according to type: {\it black:} SNIa; {\it blue:} zSNIa; {\it red:} pSNIa. The shaded area represents our host cut at $r<20$ mag. }
\label{fig:size}
\end{figure}

There exists a surprisingly tight correlation between host galaxy isophotal angular size and apparent magnitude (Fig. \ref{fig:size}). At $r_{host}= 20$ mag, the isophotal radius (at $\mu_r=25$ mag arcsec$^{-2}$) of hosts is typically 2 arcsec. (For reference, the surface brightness corresponding to S/N=1 in one arcsec$^2$ of the sky-noise-limited Annis et al. stacks is $\mu_r \simeq 27.3$ mag arcsec$^{-2}$). Two component fits at $r>20$ mag become noisy, due to the smaller number of resolution elements in the image; this is especially true for bulges. The noisiness of the fits at $r > 20$ mag is also due to the lower signal-to-noise ratio of the data: for \gimtd, the typical error in fitted bulge to total light ratio $B/T$ at $r_{host}>20$ mag is $\sigma_{B/T} > 0.07$. Furthermore, below $r_{host}=20$ mag we observe poorer consistency between the \gimtd\ and \imfit\ results. 

For the above reasons we use only those SN hosts with $r \le 20$ mag. (In \S\S\ref{sec:disks}--\ref{sec:bulge+disk} we also consider the effects of limiting the host sample to $r\le 19$ mag.) Of the 657 host galaxies with $r \le 20$ mag and $\rho_{25}<1.959$ that were fitted by \gimtd, 28 could not be fitted with \imfit. A further 29 objects were flagged by visual inspection because their fits showed strong residuals, but most of these were eliminated by other cuts on the data (all but 5 for \gimtd, {9} for \imfit). 

Another cut that can be made is on SN maximum brightness, which is related to completeness. The primary sample uses $r_{SN}^{max} < 22$ mag, which corresponds to S/N$ \simeq  7$ for a point source detection in the Stripe 82 individual scans (Dilday et al. 2008); fainter that this magnitude it is clear that the SN\,Ia data are incomplete. (The effects of completeness are addressed in more detail in Appendix \ref{subsec:completeness}.) The number of objects left after this cut is 614 (576) for \gimtd\ (\imfit). We have experimented with brighter values of limiting $r_{SN}^{max}$, and discuss these results in \S\S \ref{sec:disks}--\ref{sec:bulge+disk} below. 

A magnitude cut in $r_{SN}^{max}$ corresponds to a redshift cut, because SNe\,Ia are approximately standard candles. At a given host apparent brightness, higher redshift corresponds to higher intrinsic luminosity and larger physical size, so that the angular diameter, and hence the quality of the fit, is not too strongly redshift dependent. 

Our primary sample contains all SN\,Ia types (SNIa, zSNIa, pSNIa). Eliminating the pSNIa class (photometrically classified SNe, photometric redshifts) removes 30 objects, and makes no difference to any of the results. We eliminate the noisiest fits using the cuts  $\chi^2_r < 2$  and $\sigma_{B/T} < 0.07$ (where $\chi^2_r$ is the reduced $\chi^2$ of the fit, and $\sigma_{B/T}$ is the $1\sigma$ error in $B/T$), leaving a total of 412 (433) objects for the \gimtd\ (\imfit) fits. 

Table \ref{Tbl1} summarizes the constraints that define the primary sample of objects.

\begin{table}
\caption{Definition of the Primary Sample}
 {\renewcommand{\arraystretch}{1.0}  
\begin{tabular}{lcc}
\hline
 Sample & GIM2D\tablenotemark{a} & imfit\tablenotemark{a} \\
 \hline
 \hline
 Sako 2018\tablenotemark{b} & \multicolumn{2}{c}{1910} \\
 $r_{host}<20$, $\rho_{25}<1.959$ & \multicolumn{2}{c}{685} \\
fit converges & 657 & 629\tablenotemark{c} \\
$r_{SN}^{max}<22 $\tablenotemark{d}  & 614 & 576 \\
$\sigma_{B/T}<0.07$ & 527 & 539 \\
$\chi^2_r<2$ & 417 & 442 \\
Primary sample\tablenotemark{e} & 412 & 433 \\ 
 \hline
 \hline
 \end{tabular}  }
\begin{flushleft} 
{\tablenotetext{a}{Number of SN\,Ia, obtained by combining the constraint on a given line with all previous lines. See \S \ref{subsec:sample}.}
\tablenotetext{b} {Sako (2018) sample, SNIa + zSNIa + pSNIa
but not SNIa?, with hosts as identified in this paper. }
\tablenotetext{c}{\imfit\ was run only on those objects fitted by \gimtd.}
\tablenotetext{d}{6 objects do not have redshift information, and have been excluded.}
\tablenotetext{e}{After a visual inspection of the residuals.} }
\end{flushleft}
\label{Tbl1}
 \end{table}
\normalsize 

Finally we make a small adjustment to the fits to eliminate poorly constrained components: strongly disk-dominated systems ($B/T < 0.05$) are transformed into pure disks ($B/T=0$), and strongly bulge-dominated systems ($B/T > 0.95$) are transformed into pure bulges ($B/T=1$). 

\section {A First Look at Supernovae and Host Galaxy Light}
\label{sec:firstlook}

Before we turn to the \linclacc\ analysis, we first examine what can be said about supernovae and host galaxy light distributions as a function of angular radius.

\begin{figure}
\epsscale{1.0}
\plotone{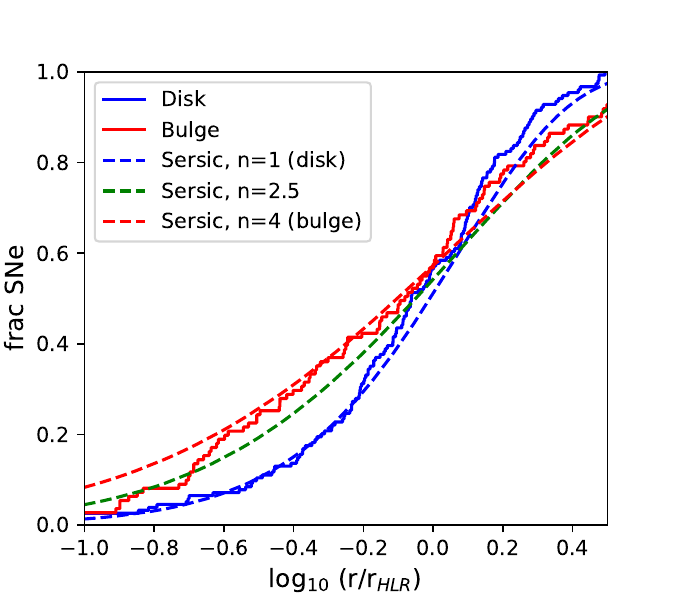}
\caption { Cumulative distribution of the radial offset of SNe\,Ia divided by the fitted half light radius of their host galaxies. {\it Red solid line:} bulge dominated objects ($B/T > 0.75$); {\it blue solid line:} disk dominated objects ($B/T < 0.25$). This is for the primary sample of objects, and the \gimtd\ analysis. The dot-dash lines are S\'ersic models: {\it blue:} $n=1$ (disk); {\it green:} $n=2.5$; {\it red:} $n=4$ (bulge). The S\'ersic profiles have been corrected for missing light at $\rho_{25} > 1.959$, and hence do not pass through ($R/HLR=1, f=0.5$).}
\label {fig:cumhlr}
\end {figure}

To remove the effects of physical size and distance, we normalize each SN radial offset by the half-light angular radius, HLR, of its host. (This HLR is derived from the fits, and is therefore not strongly affected by seeing.)  The results are shown in Fig. \ref{fig:cumhlr} for both bulge- and disk-dominated galaxies, and are compared with 3 pure S\'ersic profiles.  The main conclusion to be drawn from Fig. \ref{fig:cumhlr} is that SNe\,Ia in bulge- and disk-dominated galaxies have very different spatial distributions. SNe\,Ia in disks roughly follow the $n=1$ S\'ersic profile for the light; SNe\,Ia in bulges follow a more centrally concentrated S\'ersic profile, with $n\simeq 2.5-3$ (though $n=4$ is not as good a match to the data).  

Nevertheless the normalization by HLR (Fig. \ref{fig:cumhlr}) has problems: the analysis does not include the effects of galaxy ellipticity, or of multiple components. We therefore turn to the \linclacc\  analysis, first looking at disk-dominated galaxies.

\section {Disk Galaxies}
\label{sec:disks}

The \linclacc\  metric provides a convenient and rigorous way of comparing the distribution of supernovae with the underlying (i.e. unaffected by seeing) distribution of light in galaxies. \linclacc\  takes into account galaxy size, ellipticity, and orientation, and furthermore uses only supernovae and light with  $\rho_{25} \le 1.959$ (i.e. within the isophote at which supernovae and hosts are considered to be associated).

\begin{figure*}
\epsscale{1.1}
\plotone{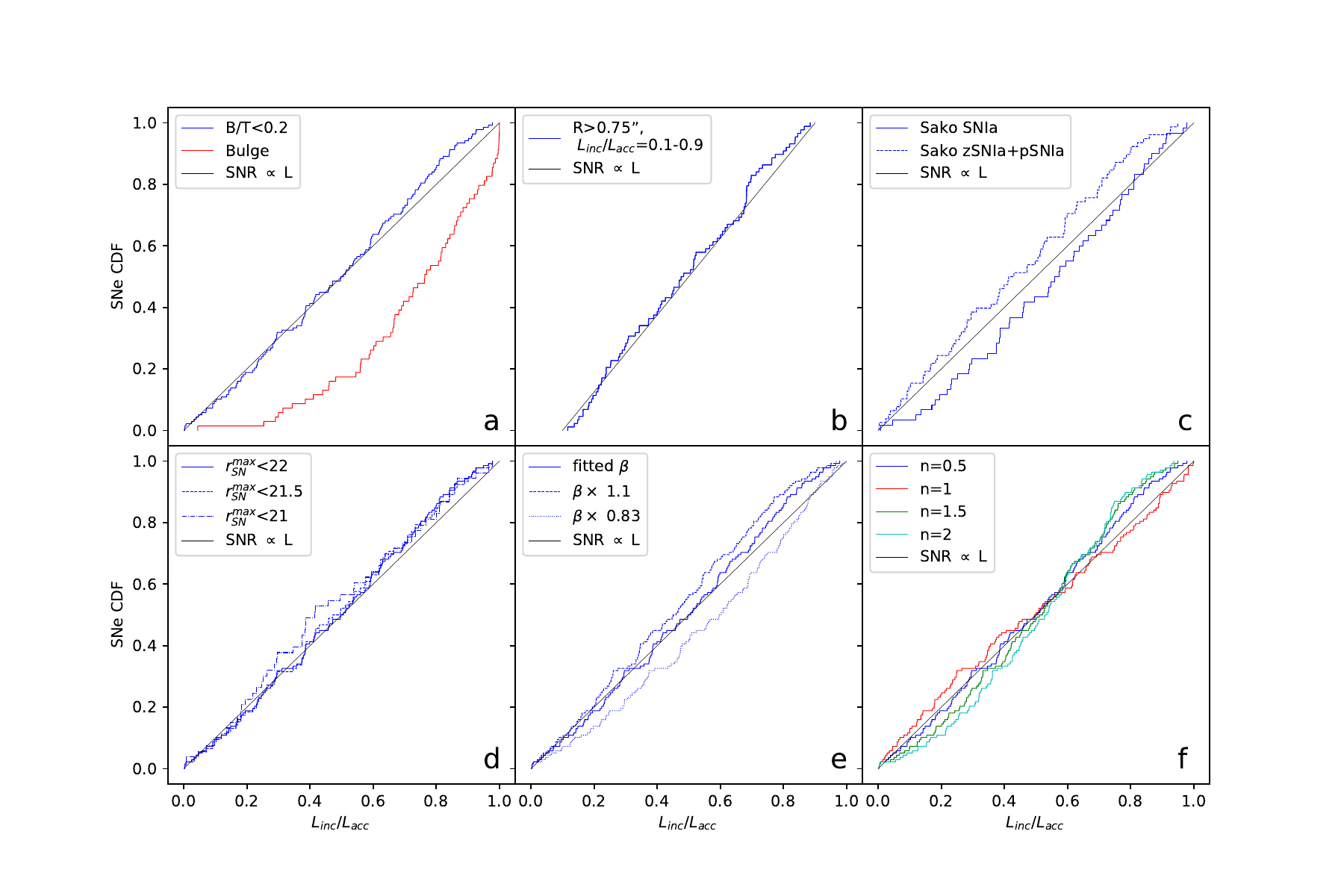}
\caption {Cumulative distribution of SNe\,Ia plotted against \linclacc, for disk-dominated galaxies ($B/T < 0.2$) in the primary sample. The diagonal line is a 1:1 relation, and corresponds to the case where the SN rate follows light. {\it Blue solid line:} \linclacc\ is calculated using the fitted disk light.
{\it (a)} Comparison with \linclacc\ calculated using fitted bulge light {\it (red line)}. 
{\it (b)} Using only SNe\,Ia outside the host galaxy core ($R > 0.75$ arcsec, $0.1 < L_{inc}/L_{acc} < 0.9$).
{\it (c)} Sako class SNIa objects ({\it solid line}) compared with zSNIa and pSNIa objects ({\it dashed line}).
{\it (d)} $r_{SN}^{max} < 22$  ({\it solid line}),  21.5 ({\it dashed line}), and 21 ({\it dot-dash line}) . 
{\it (e)} Effects of rescaling disk scale lengths $\beta$. {\it Solid line:} fitted $\beta$; {\it dashed line:} $\beta \times 1.1$; {\it dotted line:} $\beta \times 0.83$.
{\it (f)} Effects of modifying disk S\'ersic index $n$: {\it red:} $n=0.5$; {\it blue:} $n=1$; {\it green:} $n=1.5$; {\it turquoise:} $n=2$. }
    \label{fig:disk}
\end{figure*}

Fig. \ref{fig:cumhlr} shows that a single-component S\'ersic model with $n=1$ is a reasonable description of disk-dominated galaxies. We  use a  cutoff $B/T < 0.2$ on the primary sample, and examine the distribution of \linclacc\ for these disks  in Fig. \ref{fig:disk}a. The SN radial distribution matches the light distribution in the disk. Under the null hypothesis that the SN radial distribution and disk light distribution are drawn from the same parent distribution, the $p$-values for the Kolmogorov-Smirnov and Anderson-Darling tests are $p_{KS}=0.482, p_{AD}=0.415$ -- meaning that the null hypothesis is acceptable. Not surprisingly, SNe\,Ia in disk galaxies do not follow the bulge light distribution. The light distributions for this figure are from \gimtd\ fits; the results using \imfit\ fits are very similar.  The same result is obtained using a single component model fit  restricting the S\'ersic index to $n < 1.5$; SNe follow light for such a selection of objects.

Could the use of an upper limit $B/T<0.2$, with its small but not non-negligible bulge contribution, have affected these results? To check this, we tried two other limits on $B/T$ ($<0.1, <0.05$) to define the disk-dominated sample. The results are very similar to those for $B/T<0.2$.

We also try 2 techniques for removing the effects of residual bulge or nuclear light, while reducing the problem of incompleteness near the cores of galaxies ({\it if} it exists -- e.g. Appendix \ref{subsec:completeness}). We analyze supernovae and host light only in the range $0.1 < L_{inc}/L_{acc} < 0.9$ (where the upper limit also excludes a few SNe at very large radii); and we consider only SNe with $R > 0.75$ arcsec (i.e. outside the seeing disk). The result of applying both of these constraints is shown in Fig. \ref{fig:disk}b. Again, the match to the ``SNe follow light" hypothesis is excellent ($p_{KS} = 0.642$, $p_{AD} = 0.643$). Using either of the above 2 constraints alone likewise produces excellent agreement, for both \gimtd\ and \imfit\ fits.

In Fig. \ref{fig:disk}c we compare the radial distributions (in disk dominated host galaxies, $B/T<0.2$) of Sako class SNIa objects (spectroscopically confirmed, $N=60$) with Sako's zSNIa and pSNIa classes (photometric supernovae, $N=76$, of which 67 are zSNIa). There is a difference between these two samples, in the sense that the spectroscopically-confirmed sample is underrepresented at small separations. This is a clear and expected (see Appendix \ref{subsec:completeness}) signature of incompleteness at small separations in the Sako {\it spectroscopically-confirmed} (SNIa) sample.

{Now consider the full SNIa+zSNIa+pSNIa sample.} If there were a significant radial variation in SN completeness, then noticeable differences would be expected in the \linclacc\  plots for different values of the limiting SN magnitude at maximum light, $r_{SN}^{max}$. In Fig. \ref{fig:disk}d we vary the limiting $r_{SN}^{max}$ for our primary sample of disk galaxies ($B/T < 0.2$). The results for $r_{SN}^{max} < 22$, $<21.5$ and $<21$ mag are, either from KS and AD tests relative to the expected 1:1 relation, or from KS 2 sample tests, statistically indistinguishable from one another or from the null hypothesis (probabilities 40--90\%, using either GIM2D or imfit data). These tests therefore show no evidence for any radial variation in SN\,Ia completeness, and furthermore all support the null hypothesis that ``SNe trace light" in disk galaxies.

Radial gradients are known to exist in the mean ages \citep{Munoz-Mateos2007,Casasola2017} and metallicities \citep{vanderKruit2011} of stellar populations in galaxy disks. Age is known to affect the rate of SNe\,Ia through the delay-time distribution (DTD - e.g. \citealt{Heringer2019}), and metallicity may also affect the SN\,Ia rate (e.g. \citealt{Kistler2013}). One might therefore expect a small additional gradient in the SN radial distribution relative to the distribution of galaxy light. This could result in a scale length (or effective radius) for the supernova radial distribution different from that of the light.

To test this hypothesis, we have tried recalculating \linclacc\  using several different disk scale lengths. Fig. \ref{fig:disk}e shows the results of scaling all \gimtd\ disk scale lengths by a factor of 1.1$\times$ and 0.83$\times$ (using the approximation in Appendix \ref{subsec:approxlinclacc}). These models differ from the canonical (unscaled) model in Fig. \ref{fig:disk}a at the 95\% confidence level: there is only a $\sim$5\% KS or AD probability that these models agree with the ``SNe follow light" hypothesis. We therefore conclude that, at the 95\% confidence level, the scale length of the SN distribution is the same as that of disk light to within about 15\%. The same conclusion is drawn using \imfit\ data.

An alternate approach is to recalculate \linclacc\  with different values of S\'ersic $n_{disk}$ (again using the approximation in Appendix \ref{subsec:approxlinclacc}). It can be seen from Fig. \ref{fig:disk}f that $n=0.5$ and $n=1$ ($p_{KS}, p_{AD} > 0.40$) match the data best, in agreement with the $n=1$ exponential disk that was used in the fitting process for the light. Increasing $n$ to 1.5 results in $p_{KS}=0.122, p_{AD}=0.058$; $n=2$ can be rejected at the 99\% confidence level if SNe trace light. {In other words,} there is no evidence supporting the hypothesis that the radial distribution of SNe\,Ia follow a different S\'ersic profile from that of the disk, as might have been expected had gradients in metallicity and stellar age affected SN rates.

Finally, we consider the effects of using a sample of disk-dominated host galaxies with $r<19$ mag. The results are almost indistinguishable from those for the Primary Sample with $r<20$.

\section {Bulges and Ellipticals}
\label{sec:bulges}

It is more difficult to perform an \linclacc\  analysis for galaxies with prominent bulge components (and ellipticals). The main problem is resolution: at the typical redshifts of the SDSS SN Survey, bulges of hosts are only marginally resolved (e.g. \S\ref{subsec:hostfitting}). SN incompleteness may also be a problem
in the inner regions of bulges because of their higher surface brightness: they are both noisier due to photon noise, and also are more likely to have artifacts and substructure in the subtracted images which are used for SN detection. {(Evidence of incompleteness for inner bulge SNe may be visible in Fig. \ref{fig:cumhlr},  though the overall completeness of the survey is high \citep[see also Appendix \ref{subsec:completeness}]{dild10b}.)}

\begin{figure*}
    \epsscale{1.1}
    \plotone{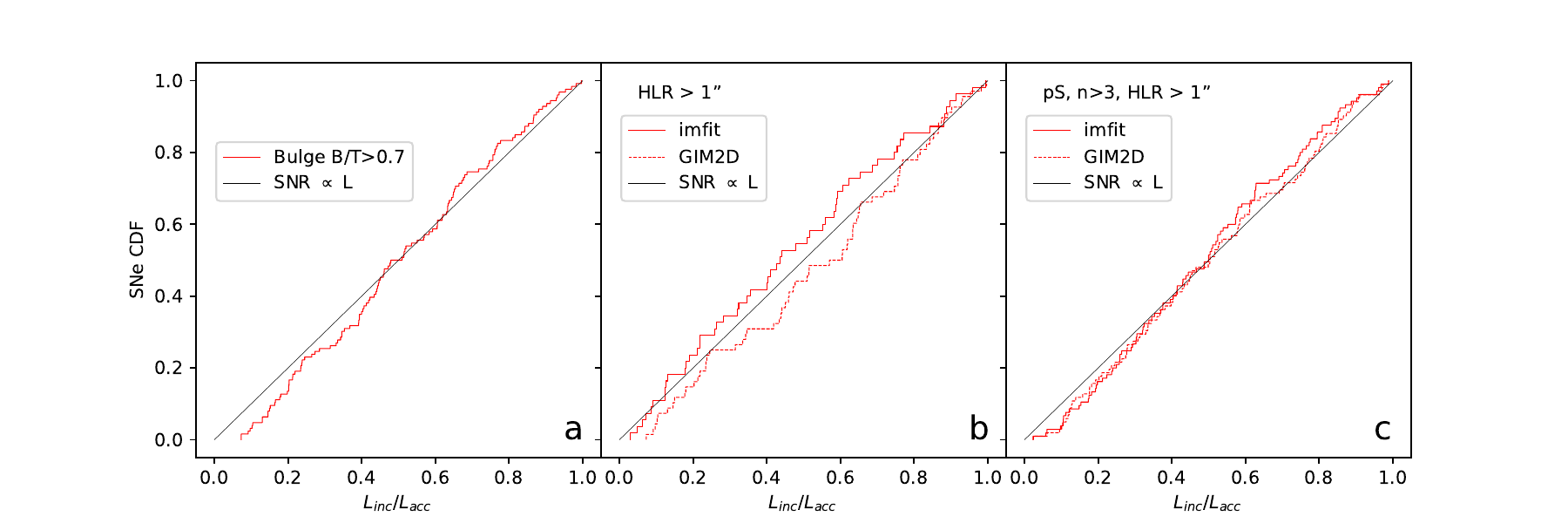}
   \caption{\linclacc\  distribution for bulge dominated galaxies ($B/T>0.7$) in the primary sample ($r<20$) {\it (a)} \gimtd\ bulge fits. {\it (b) } $HLR > 1$ arcsec. {\it Red solid line:} \imfit\ fits; {\it red dashed line:} \gimtd\ fits. {\it (c)} Single component (pure S\'ersic model), $n > 3, HLR > 1$ arcsec. {\it Red solid line:} \imfit\ fits; {\it red dashed line:} \gimtd\ fits.
    }
    \label{fig:bulge}
\end{figure*}

{
With these caveats in mind, we now look at \linclacc\  for bulge-dominated galaxies with $B/T > 0.7$ (Fig. \ref{fig:bulge}a). $p_{KS}$ and $p_{AD}$ are lower for these galaxies ($\la 0.1$). 
However, if the bulge analysis is compromised by resolution effects, then it makes sense to restrict the analysis to the largest objects, which are better resolved. The constraint $HLR \equiv r_{eff} > 1$ arcsec (i.e. including only the largest 50\% of bulges, corresponding to $err(r_{eff})/r_{eff} \la 0.1$)  improves the  probabilities (\gimtd: $p_{KS}$, $p_{AD}=$ 0.348, 0.356; \imfit: $p_{KS}$, $p_{AD}=$ 0.550, 0.516). This is shown in Fig. \ref{fig:bulge}b; results are similar for $HLR > 1.2$ arcsec.


What about the single component models? Looking only at the objects with S\'ersic $n>3$, the KS and AD probabilities are $\la 0.10$. However, including a constraint $HLR > 1$ arcsec improves the agreement with the SNe traces light hypothesis (\gimtd: $p_{KS}$, $p_{AD}=$ 0.662, 0.428; \imfit: 0.719, 0.463). This is shown in Fig. \ref{fig:bulge}c.

The KS and AD probabilities can also be raised to $>0.7$ by restricting the analysis to $0.1 < L_{inc}/L_{acc} < 0.9 (0.7)$ for \gimtd\ (\imfit). This is probably a signature of misfitting of the  $r_{eff}$ value, presumably due to resolution effects, and possibly also incompleteness in the inner regions. It may also be due to SNe being scattered out of the inner regions due to errors in the coordinates (Appendix \ref{subsec:xyerr}). Changing the value of $n_{bulge}$ does not improve the agreement between \gimtd\ and \imfit; nor does adding in the effects of the (relatively small amount of) disk light. For single component models with S\'ersic index $n>3$, there is no statistically significant difference between the results for $r<19$ and $r<20$ (2 sample KS test p-value 0.66, or 0.79 if the sample is restricted to $0.1 \le$ \linclacc\ $\le 0.9$). For two component models with $B/T > 0.7$, a similar result is obtained if the samples are restricted to $HLR > 1$ arcsec or SNe with $\Delta\theta > 0.75$ arcsec (2 sample KS p $\ge 0.80$).

We conclude that the evidence from the largest 50\% ($HLR > 1$ arcsec) of bulge-dominated hosts is consistent with the hypothesis that Type Ia supernovae and galaxy light have the same spatial distribution. 
}

\section{Galaxies with Prominent Bulges and Disks}
\label{sec:bulge+disk}

In this section we discuss objects intermediate between disk-dominated (\S\ref{sec:disks}) and bulge-dominated (\S\ref{sec:bulges}) galaxies. These objects have $0.2 < B/T < 0.7$. 

\begin{figure*}
  \epsscale{1.1}
    \plotone{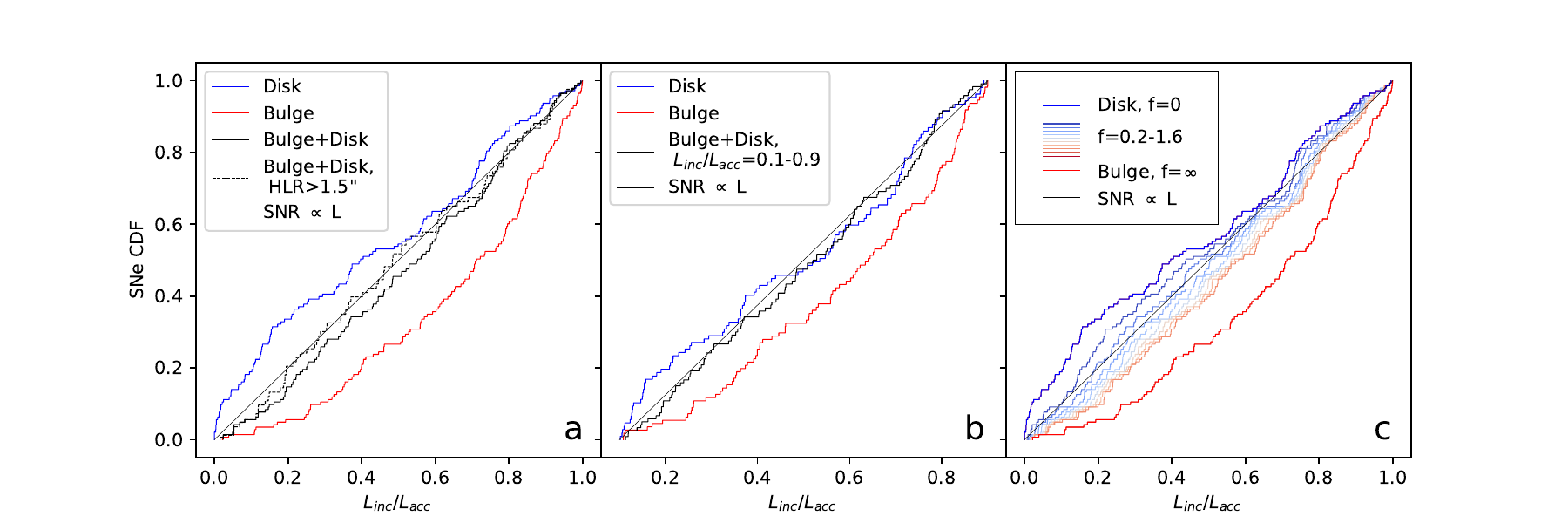}
    \caption{{\it (a)} GIM2D analysis for objects with $0.2 < B/T < 0.7$, with \linclacc\ calculated using  bulge light ({\it red}), disk light ({\it blue}), and total light ({\it black}). The black dashed line uses only objects with HLR $> 1.5$ arcsec.{\it (b)} \linclacc\ calculated over the range $0.1<$\linclacc$<0.9$, i.e. ignoring the innermost and outermost SNe. {\it (c)} Bulge SN rate scaled by $0.2-1.6\times$, in steps of 0.2.}
    \label{fig:bulge+disk}
\end{figure*}

The \gimtd\ analysis of these objects is shown in Fig. \ref{fig:bulge+disk}a. This figure shows that \linclacc\ calculated from bulge or disk light alone does not match the 1:1 relation; in other words, supernovae must occur in both components. On the other hand, \linclacc\ measured for total light provides a somewhat better match to the SN distribution ($p_{KS}$, $p_{AD}= 0.399, 0.228$). The agreement is substantially worse for the \imfit\ analysis ($p_{KS}$, $p_{AD}= 0.069, 0.059$), as was also the case for objects dominated by bulge light (\S\ref{sec:bulges}).

We have tested whether the host magnitude limit affects these results for hosts with intermediate B/T. A KS two sample test shows that results for hosts with $r<19$ are indistinguishable from the primary sample ($r<20$).

The agreement between \linclacc\ and cumulative SN numbers for $0.2 < B/T < 0.7$ can be improved in at least three ways, two of which we saw in bulge-dominated galaxies (\S\ref{sec:bulges}). First, the analysis can be restricted to $0.1 < $\linclacc$ < 0.9$, excluding SNe in the inner and outer regions. The result of this is shown in Fig. \ref{fig:bulge+disk}b; the total light fit is much better (\gimtd: $p_{KS}, p_{AD}=  0.930, 0.677$, \imfit: $p_{KS}$, $p_{AD}= 0.420, 0.389$). Second, the analysis can be restricted only to the largest objects (Fig. \ref{fig:bulge+disk}a). For objects with HLR $>$1 ($>$1.5) arcsec, the \gimtd\ probabilities are $p_{KS}, p_{AD}=$ 0.684, 0.419 (0.966, 0.934). Third, the bulge SN rate per unit light can be decreased relative to the disk rate. We defer a discussion of this to \S \ref{sec:discussion}.

\section {Discussion}
\label{sec:discussion}

\subsection{Do SNeIa Follow Light?}

As discussed in \S 1, for plausible delay-time distributions, the SN\,Ia rate per unit $r$-band luminosity, $SNR/L_r$, is expected to be more or less constant over a wide range of stellar population color, metallicity, and star formation history \citep{Heringer2016,Heringer2019}. It is therefore not surprising that the rate of SNe\,Ia and light track each other closely in disks (\S\ref{sec:disks}). 

The situation with bulges is more complex. As already noted, at the typical redshifts of the SDSS SN Survey, bulges of hosts are only marginally resolved, so the bulge parameters $r_{eff}$, $\mu_{eff}$, and $b/a$ are poorly constrained, especially with an admixture of disk light. Furthermore, nuclear light may contaminate the bulge fitting, resulting in an underestimate of $r_{eff}$. Our use of \gimtd\ and \imfit\ requires that bulges be fitted with a fixed S\'ersic index $n=4$, because of the small number of pixels in the bulge; but real bulges and ellipticals actually possess a wide range of $n$ values (Appendix \ref{subsec:bulgen}). Finally, astrometric errors may scatter SNe out of the region \linclacc $\la 0.1$ in bulges (Appendix \ref{subsec:xyerr}).

In \S\ref{sec:bulges} we noted that, for bulge-dominated host galaxies, bulge light and SNe\,Ia followed the same profile if the sample of host galaxies is restricted to the largest, best-resolved $\sim 50$\% of objects (HLR $\ga 1$ arcsec). This is true for both the 2-component and 1-component models. 


What about the objects which are intermediate between bulge-dominated and disk dominated ($0.2 < B/T < 0.7$)? Based on the above discussion, one might have expected SNe to follow total light. However, in \S\ref{sec:bulge+disk} we noted a lack of SNe in the inner regions relative to the total light \linclacc\ profile. This discrepancy is ameliorated if the sample of hosts is restricted to HLR $\ga 1$ arcsec, or if the sample of SNe is restricted to $0.1 <$ \linclacc $< 0.9$.

Another explanation is possible: the rate of SNe\,Ia per unit r-band luminosity, $SNR/L_r$, in bulges may be different from that in disks, and would hence result in a lower than expected number of supernovae in the inner regions of galaxies relative to the disk rate. We have already discussed the fact that this is unlikely for a delay-time distribution DTD $\sim t^{-1}$. However given the relatively large color difference between bulges and disks (typically $\Delta(g-r) \simeq 0.2$),
a significant difference between bulge and disk $SNR/L_r$ might be expected for DTD$\simeq t^{-1.5}$ or steeper, or for a DTD with a cutoff (cf. \citet{Heringer2019}, Fig. 7).

Fig. \ref{fig:bulge+disk}c shows the result of scaling the bulge SN rate relative to the disk rate, for \linclacc\ calculated using total light. (This plot is computed using the approximation in Eq. \ref{eq:linclaccscaling}.)
As expected, for small bulge rate scaling factor $f$, \linclacc(tot) approaches that for disk light; for larger $f$, \linclacc(tot) approaches the bulge results. The best scaling factor is $f\simeq0.5\pm0.2$, where $p_{KS}, p_{AD} = 0.858, 0.802$ at $f=0.5$, and the quoted error range corresponds to $p \simeq 0.5$. This is for the \gimtd\ analysis; the results for \imfit\ are similar. 

Referring to Fig. 7 of \citet{Heringer2019}, if bulges are old, then this result is consistent with a DTD $\sim t^{-1.5}$, or a power-law DTD with a cutoff (a steeper power-law slope) at $t>1-2$ Gyr. However, as already mentioned, this result can also be explained as due to resolution effects; the correct explanation must await host (and supernova) observations with better resolution.

\subsection{Astrophysical Effects}

Here we consider how the properties of supernovae, or of their host galaxies, affect the conclusions regarding the spatial distribution of supernovae. First we examine the effects of SN light curve width or stretch ($s$ or $x1$) and color at maximum light ($c$). These photometric parameters are used by the SALT2 model to empirically determine the absolute magnitude of a SN\,Ia  \citep{Guy2007,Sako18}; although both parameters are required by SALT2, $c$ is significantly more sensitive in determining absolute magnitude. 

{There exists a great deal of work on the properties of SN\,Ia host galaxies,  and especially on how these properties relate to SN\,Ia light curves (e.g. \cite{murakami2023a} and references therein). Earlier work by, for example, \cite{sullivan2006} (see also \citealt{Maoz14} and references therein) showed that} the stretch parameter $s$ correlates with host galaxy sSFR: SNe in galaxies with high sSFR tend to have broad light curves with stretch $s>1$, whereas elliptical and bulge-dominated spiral hosts with low or zero sSFR have an $s$ distribution dominated by SNe with $s<1$. Furthermore, SNe\,Ia with $s<1$ appear to be associated with galaxy mass, whereas $s>1$ SNe\,Ia appear to be prompt objects associated with star formation. These conclusions apply to the integrated properties of galaxies; the question arises whether they may also apply {\it within} galaxies. In other words, are the spatial distributions of $s<1$ and $s>1$ SNe (or $c < 0$ and $c > 0$ SNe) different?

\begin{figure}
    \plotone{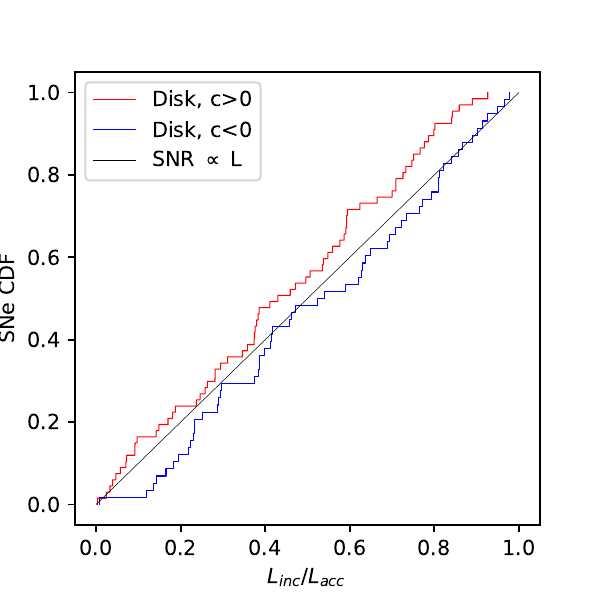}
   \caption{Disk-dominated galaxies ($B/T < 0.2$) analyzed by \gimtd. {\it Red:} supernovae with color at maximum light $c>0$; {\it Blue:} supernovae with $c<0$.}
    \label{fig:disk.color}
\end{figure}

Fig. \ref{fig:disk.color} compares the \gimtd\ \linclacc\ distributions for disk galaxies ($B/T < 0.2$) hosting red ($c>0$) and blue ($c<0$) SNe\,Ia. Formally these two distributions are in acceptable agreement ($p=0.36$ from a 2-sample KS test). Of interest, however, is the lack of blue SNe\,Ia in the inner regions of disk-dominated galaxies: of the 14 SNe\,Ia at \linclacc$<0.1$, only 2 have $c<0$. Under the null hypothesis that red and blue SNe possess equal likelihood (which is suggested by the overall numbers of supernovae in each category), the probability of finding $\le 2$ red or blue objects is only $p=0.013$. Similar results are found with \imfit. A much weaker effect exists for the blue SNe\,Ia in hosts with mixed morphology ($0.2 < B/T < 0.7$): the null hypothesis that SNe trace total light in these hosts has $p_{KS}, p_{AD}=  0.225, 0.119$), with no statistically significant difference between the red and blue subsamples. No color separation is seen for the bulge-dominated hosts ($B/T > 0.7$).

Could the SN color effect in disks be due to incompleteness in the cores of galaxies? Typical blue SNe\,Ia are 0.5-1 mag {\it brighter} than typical red SNe\,Ia, in the opposite direction of what would be expected from incompleteness.

This anomaly could be due to a gradient in some property in these galaxies, coupled with a dependence of $c$ on this property. For example, if SN $c$ depends on mean stellar age $\langle\tau\rangle$, and $\langle\tau\rangle$ is different in the cores of disk galaxies compared to their outer regions, then the observed effect (missing blue SNe in the core) could be a consequence. We have examined the \linclacc\ distributions for $c<0$ and $c>0$ SNe\,Ia for hosts with even smaller amounts of bulge light ($B/T < 0.1$ and $B/T < 0.05$). The results are almost identical, although with lower significance because of the smaller numbers of objects involved. We conclude that, if $c$ depends on SN age, it is not the age difference between bulges and disks that is causing this effect.

Finally, the presence of significant amounts of dust in the central regions of disk galaxies could obviously redden core SNe\,Ia. 

No significant differences are seen in the radial distribution of SNe\,Ia that depend on the stretch parameter $s$. However, as noted by \cite{sullivan2006}, there are very significant differences in the numbers of low stretch ($s<1: n=96$), and high stretch ($s>1: n=29$) SNe\,Ia in bulge dominated ($B/T>0.7$) galaxies. This is  a $\sim4\sigma$ difference compared to the equal numbers of $s<1$ and $s>1$ objects observed in all SDSS host galaxies.

\begin{figure}
    \plotone{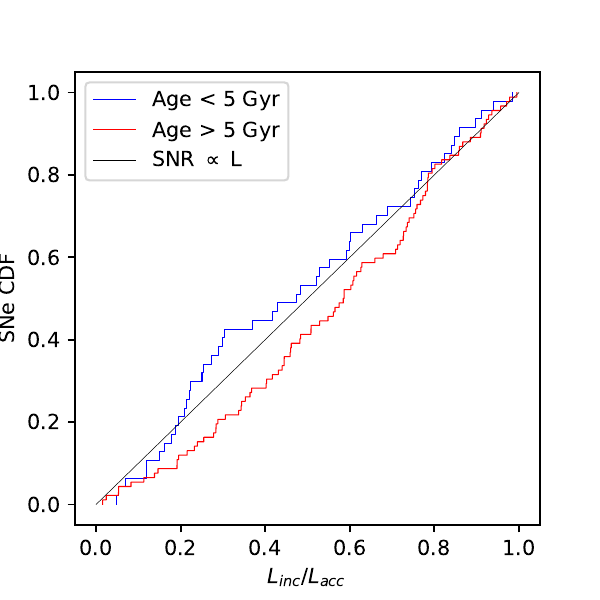}
    \caption{Intermediate B/T host galaxies ($0.2 < B/T < 0.7$) analyzed by \gimtd. {\it Red:} hosts with mean age $>5$ Gyr; {\it Blue:} hosts with mean age $<5$ Gyr. The age is as determined by FSPS \citep[][column 124 of Table 1]{Sako18}.}
    \label{fig:bulge+disk.agegyr}
\end{figure}

Do our conclusions on the radial distribution of SNe\,Ia depend on the properties of the host galaxies? We have examined the following host galaxy properties in the \citet{Sako18} database: absolute magnitude, mass, $g-r$ color, mean age, specific star formation rate, and number of companions. (Many of these properties are correlated.) For each property, we have divided the SNe into two roughly equal-sized samples, and have compared the \linclacc\ distributions for bulge-dominated, disk-dominated, and intermediate hosts, using a 2 sample KS or AD test. No significant differences were seen, except in one case: intermediate B/T ($0.2 < B/T < 0.7$) hosts divided according to whether the mean stellar population age is $<5$ Gyr or $>5$ Gyr (Fig. \ref{fig:bulge+disk.agegyr}). Here we see a strong difference between the two samples, in the sense that older intermediate B/T galaxies have disproportionately more SNe\,Ia in their outer regions: from a KS 2 sample test the probability that these 2 samples are drawn from the same distribution is only 6\%. 

{However, this result should be viewed with some caution. Determining star formation history  and mean age is difficult from broadband photometry alone \citep{conroy2013}, even with the addition of UV and IR fluxes to SDSS $ugriz$  \citep{Gupta2011}. In addition, galaxy age is observed to correlate with $B/T$ in the SDSS host sample; this is expected  because bulges are older than  disks. Dividing the data by $B/T$ rather than by age in fact produces a plot similar to Fig. \ref{fig:bulge+disk.agegyr}. Thus it is likely that $B/T$ is the driver of the age effect. }


\section{Conclusions}\label{sec:conclusions}

We have used the SDSS SN Survey to study the radial distribution of SNe\,Ia in their host galaxies. Each host galaxy has been decomposed into a seeing-convolved bulge plus disk model, and the distribution of SNe has been compared with the distribution of model light enclosed by an isophote passing through each SN (\S \ref{subsec:linclacc}).

We find that the rate of SNe\,Ia in disk-dominated galaxies is proportional to the $r$-band surface brightness of the disk. Not surprisingly, a S\'ersic index $n=1$ (i.e. an exponential disk) provides the best fit for the distribution of supernovae. There is no evidence that the disk scalelength (or equivalently $r_{eff}$) is different for the radial distributions of SN\,Ia and light.

The situation with bulge-dominated and elliptical galaxies is less clear because of resolution effects. Nevertheless, the evidence from the larger $B/T>0.7$ galaxies also shows that the SN\,Ia rate is proportional to light.

For host galaxies with intermediate $B/T$  ($0.2 \le B/T \le 0.7$), it is clear that SNe\,Ia occur in both the disk and bulge components. There is a lack of SNe\,Ia in the inner regions of these galaxies. This could be due to incompleteness or resolution effects, as was the case for bulge-dominated galaxies. It could also be due to a smaller SN rate per unit light in bulges (as compared with disks). 

There is evidence for missing blue (SALT2 $c<0$) Type Ia supernovae in the cores of disk-dominated galaxies. This is possibly due to the effects of central dust. 

One major limitation in this work is the lack of resolution of bulges and ellipticals at redshifts $z \simeq 0.2-0.4$ in SDSS data. In the future we plan to avoid  this limitation by using HST imaging for the host light fitting, and by redoing the analysis on a lower redshift transient survey such as the Zwicky Transient Facility (ZTF)\footnote{\url{https://www.ztf.caltech.edu/}}.

\acknowledgments
\noindent Special thanks to Luc Simard for his assistance with \gimtd\ in the early stages of this project, to Sara Ellison and Trevor Mendel for valuable help in navigating the SDSS databases, to Peter Erwin for assistance and advice in setting up \imfit, to Masao Sako for information in advance of publication, to Peter Garnavich for helpful comments, and to the grad students of Nanjing University for supplying hardware support. CB gratefully acknowledges support from the Natural Sciences and Engineering Research Council of Canada under their postdoctoral fellowship program and the Forrest Research Foundation. {We are grateful to an anonymous referee for helpful  comments.}. 

The results reported here are based on SDSS DR7 photometric and spectroscopic catalogs and on the SDSS-II Supernova Survey. Funding for the SDSS and SDSS-II has been provided by the Alfred P. Sloan Foundation, the Participating Institutions, the National Science Foundation, the U.S. Department of Energy, the National Aeronautics and Space Administration, the Japanese Monbukagakusho, the Max Planck Society, and the Higher Education Funding Council for England. The SDSS Web Site is http://www.sdss.org/. 

The SDSS is managed by the Astrophysical Research Consortium for the Participating Institutions. The Participating Institutions are the American Museum of Natural History, Astrophysical Institute Potsdam, University of Basel, University of Cambridge, Case Western Reserve University, University of Chicago, Drexel University, Fermilab, the Institute for Advanced Study, the Japan Participation Group, Johns Hopkins University, the Joint Institute for Nuclear Astrophysics, the Kavli Institute for Particle Astrophysics and Cosmology, the Korean Scientist Group, the Chinese Academy of Sciences (LAMOST), Los Alamos National Laboratory, the Max-Planck-Institute for Astronomy (MPIA), the Max-Planck-Institute for Astrophysics (MPA), New Mexico State University, Ohio State University, University of Pittsburgh, University of Portsmouth, Princeton University, the United States Naval Observatory, and the University of Washington.

\vspace{5mm}
\facilities{APO, SDSS}

\software{astropy \citep{Astropy1, Astropy2, Astropy3}, numpy \citep{Numpy1}, matplotlib \citep{Matplotlib}, pgplot,
    \imfit\ \citep{Erwi15},
          GIM2D \citep{Sima02}, 
          SExtractor \citep{Bert96}
          }

\clearpage
\appendix

\section{Fitting}
\label{app:hostfitting}
\subsection{Stacked Images and PSFs}
\label{subsec:PSF}

There is considerable variation in the point spread function (PSF) over the field of the SDSS detector, and of course time variability in the PSF along each individual scan due to seeing variations. \citet{Anni14} provide a spatially interpolated PSF\footnote{This PSF is a weighted superposition of the individual PSFs that went into the stack.} for all locations in the S82 stacks, and this was used for both the \gimtd\  and  \imfit\ host galaxy fits. The median full width at half maximum (FWHM) for the PSF's of the full sample of hosts with $r < 20$ is 1.0 arcsec for the stacked images. 

For point source detection, the stacked images have a $S/N=5$ limiting magnitude $r_{lim}=23.5$ mag, and are typically $>$1.5 mag deeper than the single scans. 

At $r=19$ mag the typical internal error in the fitted value of the bulge to total light ratio\footnote{$B/T$ depends on wavelength; in this paper $B/T$ always refers to the $r$ band.} $B/T$ is $\pm 0.033$ for stacked images, vs. $\pm 0.23$ for single scans (for a galaxy with $B/T=0.5$). Nevertheless the median reduced $\chi^2$ value for the fits is higher for the stacks: at $r=19$ mag, $\langle\chi_\nu^2\rangle \simeq 1.5$ for stacks, vs. $\sim 1$ for single scans. This may be due to the fact that the stacks are so much deeper ($>1.5$ mag) than single images;  substructure not fitted by the simple bulge+disk models has a much greater effect in higher S/N images.

\subsection {GIM2D Light Profile Fitting}
\label{subsec:gim2d}

For the \gimtd\ fits, we use the publicly available Stripe 82 fits \citep{Bott19}, augmented with additional fits for fainter galaxies using the same techniques. \citet{Bott19} adopted the same fitting approach as \citet{Sima11}, who provide an extensive description of the methodology used in constructing and characterizing these catalogs. 
Table B1 in \citet{Bott19} provides a complete list of all the fitted morphological and associated photometric values returned by \gimtd\ for each galaxy. 

{\gimtd\ provides Markov Chain Monte Carlo  (MCMC) uncertainties by sampling the convergence region to obtain the posterior probability distribution in each parameter \citep{Sima11,Bott19}. The best-fitting values and associated uncertainties are estimated using the marginal distributions of each parameter. In principle these should reflect the uncertainties associated with a repeat fit under the same observing conditions. }

\subsection {Imfit Light Profile Fitting}
\label {subsec:imfit}

The \citet{Anni14} stacked images 
have a fitted sky background level removed. Nevertheless there remain local sky variations over the field, due to scattered light, flat fielding errors, etc. For the \imfit\ fits we measured the local sky surrounding each host galaxy using an elliptical annulus with shape and orientation taken from SDSS $a$ and $b$, and annular size 1.5--2.5 $\rho_{25}$ (where $\rho_{25}$ is defined in \S \ref{subsec:hostmatching}). The amount of sky that had originally been subtracted from the images (required for the \imfit\ noise model) was estimated as that sky level required to produce the observed sky noise (from $\sigma^2_{sky}=S_0/g_{eff}$, where $S_0$ is the sky level in DU, and $g_{eff}$ is the effective gain provided by \citealt{Anni14}).

Pixels contaminated by neighboring objects were removed (masked) from the fit. The mask image was constructed by running SExtractor \citep{Bert96} with $thresh=3$, and masking pixels that lay within an elliptical aperture with semi-major axis $a_{mask}=1.25 a_{Kron}$, where $a_{Kron}$ is the \citet{Kron80} semi-major axis as measured by SExtractor. A further masking of contaminants was performed manually by inspecting the residuals of a preliminary run of \imfit. About half of all host galaxies were found to have additional faint neighbours that were then masked; this manual masking made virtually no difference to the final fits.

A number of different \imfit\ fits were run for the two component (bulge + disk) models, with 3 different minimization algorithms and many different starting points. 

{To compute errors in the fitted parameters, we used the standard \imfit\ procedure of refitting 100 bootstrapped images (resampled with replacement),  starting at the previously-determined best $\chi^2$ solution. The 1$\sigma$ parameter errors were computed from the distributions of bootstrap fitted values. Unlike \gimtd, \imfit\ does not fit $B/T$; rather, it is calculated from 6 other fitted parameters. $\sigma_{B/T}$ is simply calculated by estimating $B/T$ separately for each bootstrap iteration, and using the marginal distribution of these values.}

\section{More on \linclacc}
\label{sec:morelinclacc}

\subsection{Approximations}
\label{subsec:approxlinclacc}

For the total light calculation one can also allow for a different $SNR/L_r$ in the bulge and disk by simple scaling: 

\begin{equation}
\frac{L_{inc}^{tot}}{L_{acc}^{tot}} \simeq \frac { L_{inc}^d + f L_{inc}^b} {L_{acc}^d + f L_{acc}^b}
\label{eq:linclaccscaling}
\end{equation}

\noindent
 where $tot$ refers to total light, $b$ is bulge, $d$ is disk, and $f$ is the factor by which $SNR/L_r$ in the bulge is scaled relative to that in the disk. Strictly speaking a full (time-consuming) integration is required to do this correctly, but numerical experiments show that eqn. \ref{eq:linclaccscaling} is a good approximation for most galaxies.

In the analysis that starts in \S \ref{sec:disks}, we look at the effects of varying the model scale lengths and S\'ersic indices on \linclacc. To do this exactly would be a great deal of work in recomputing the time-consuming \linclacc\  integrations. Again there is a simple intuitive approximation, as follows. 1. Calculate $\alpha_{inc}=R_{inc}/R_{eff}$ that corresponds to $L_{inc}$ for the component of interest. 2. In the same way find $\alpha_{acc}$ for $L_{acc}$. 3. Calculate the new $\alpha{'}_{inc} = \alpha_{inc}/\beta$, and the same for $\alpha{'}_{acc}$. Here $\beta$ is the amount by which the original $R_{eff}$ is scaled. 4. Recompute $L_{inc}$ and $L_{acc}$ using the new $\alpha{'}_{inc}$ and $\alpha{'}_{acc}$, and the new Sersic index $n'$. 5. Calculate the new value of \linclacc. Numerical experiments show that this approximation for perturbing models is adequate for our purposes.

\subsection {Comparison of GIM2D and imfit Fitting}
\label {subsec:comparisonfit}

\begin{figure*}
\epsscale{1.1}
\plotone{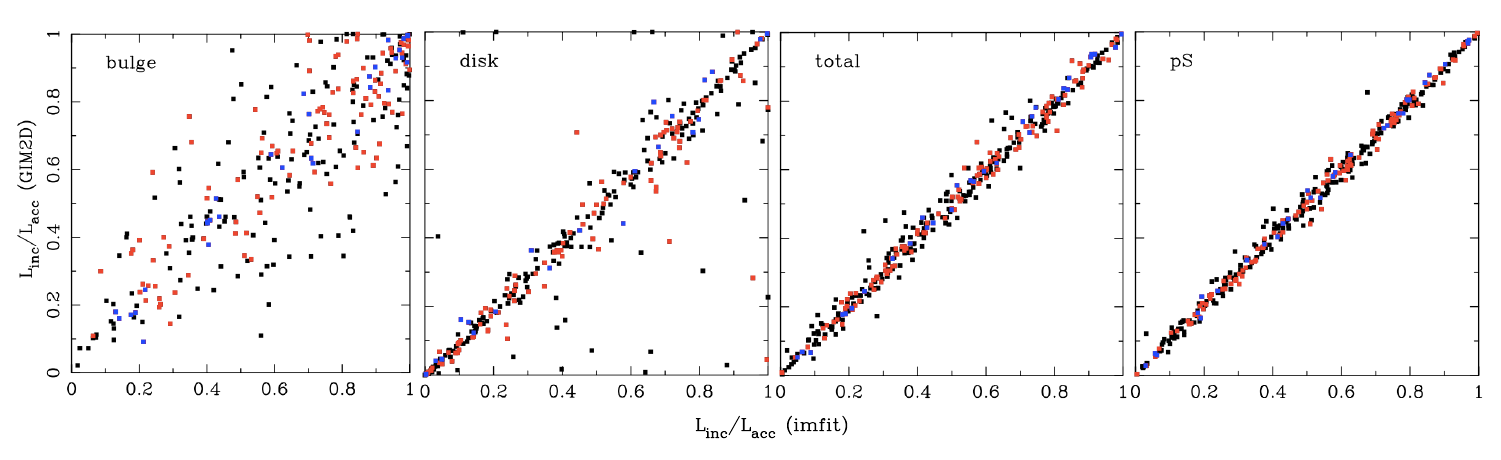}
\caption {Comparison of \linclacc\  from \gimtd\ and \imfit\ fits of the host galaxy light distribution. The first three panels are from two component (bulge plus disk) models, with \linclacc\  calculated from (left to right) fitted bulge light only, fitted disk light only, and fitted total (i.e. bulge plus disk) light. The right panel is calculated using a single component model (i.e. pure S\'ersic or pS model). The plotted points are from the Primary Sample of objects (\S \ref{subsec:sample}): {\it black} -- $r_{host} < 20$, {\it red} -- $r_{host} < 19$, {\it blue} -- $r_{host} < 18$. }
\label{fig:gimvsim}
\end{figure*}

How well do \gimtd\ and \imfit\ models agree? To answer this question, we compare computed \linclacc\  values (Fig. \ref{fig:gimvsim}), because these are the principal diagnostics used in this paper. The sample of objects is the ``Primary Sample'' discussed in \S \ref{subsec:sample}. \linclacc\  for the disk component, and also total light (the sum of the bulge and disk components) agree very well between the two programs. Not surprisingly, however, the bulge light component shows considerable scatter, because bulges are small and poorly resolved in these data (\S \ref{subsec:hostfitting}). The scatter in the fitted value of $B/T$ is $\sigma_{B/T}=0.08$, assuming each program contributes equally to the disagreement. We use this as a rough estimate of the systematic error in the fitted value of $B/T$. (For $B/T=0.5$ the statistical $1\sigma$ error reported by the fitting programs is typically about $\pm 0.07$ for $r=20$, and $\pm 0.033$ for $r=19$.)

Also shown is a comparison of \linclacc\  for the single component (pure S\'ersic) models: \gimtd\ and \imfit\ are in excellent agreement for these fits.

\section{Biases in the Profile Fitting}
\label{sec:biases}

\subsection{Completeness}
\label{subsec:completeness}

In principle the overall completeness of our SN\,Ia sample, or of our host galaxies, should have no effect on our methodology, provided there is no {\it radial gradient} of SN completeness in our host galaxies. But radial gradients in completeness must exist, at least in the sub-sample of spectroscopically confirmed SDSS supernovae (Sako class SNIa): as discussed in Sako et al (2008), fainter SNe\,Ia ($r \ga 20.5$ mag) are selected for follow-up according to a weight that depends on the amount of host galaxy light contamination. (The inclusion of photometric SNe\,Ia (classes zSNIa and pSNIa) compensates for this radial effect.)

Dilday et al. (2008) discuss SDSS SN survey detection efficiency using artificial SNe injected into the survey data. According to Dilday et al. (2010b ApJ 715), ``Analysis of these artificial SN\,Ia from the three observing seasons of the SDSS-II Supernova Survey does not show evidence for significant loss of efficiency near the cores of galaxies."

To test this, we have compared the redshift distributions and SN maximum light distributions for SNe inside and outside the seeing disk. We find that the redshift distribution of inner ($R<0.75$ arcsec) SNe is shifted $\sim 0.02$ to {\it larger} z, and the apparent magnitude distribution is shifted by $\sim 0.1$ mag to {\it fainter} magnitudes. These two effects are small, and in the opposite direction of what would be expected if the detection efficiency were lower in the central regions of host galaxies.

In the analysis (\S\S\ref{sec:disks}--\ref{sec:bulge+disk}) we try three additional tests for incompleteness in the cores of galaxies: (1) we restrict the analysis to galaxies hosting brighter supernovae; (2) we compare  \linclacc\  distributions excluding objects at small \linclacc; and (3) we consider only supernovae outside the cores of galaxies, using a host-SN angular separation restriction ($\Delta\theta > 0.75$ arcsec), and calculating \linclacc\ starting at this inner boundary. 

{Incompleteness may be more of a problem in the cores of bulge components. Because bulge cores have higher surface brightness, photon noise is greater, and image subtraction artifacts are more of a  problem. }

\subsection {Supernova Light Contamination Bias}
\label {subsec:SNcontam}

\begin{figure}
\epsscale{1.0}
\plotone{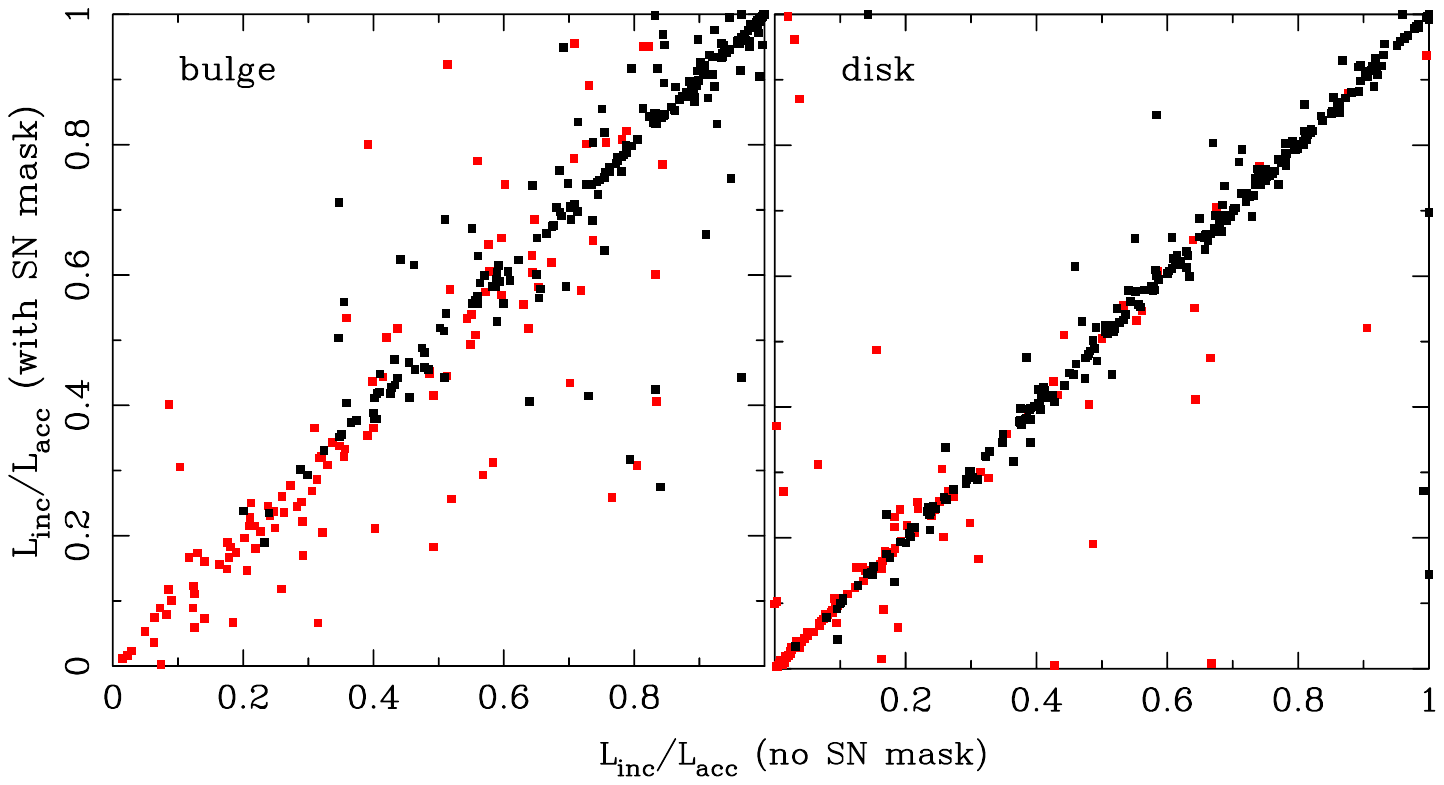}
\caption{\linclacc\  ratio for bulge ({\it left}) and disk ({\it right}) fits with SN masking vs. no SN masking. {\it Black:} $\Delta\theta_{SN} > 0.75$ arcsec; {\it red:} $\Delta\theta_{SN} < 0.75$ arcsec.}
\label{fig:llSNmask}
\end{figure}

The Annis et al. stacks were made using some epochs that included SN light. Is it possible that this SN light bled through into the stacks, and biased our galaxy fitting? To test this, we repeated the \imfit\ fitting procedure with the SN position masked for each host, using a circular mask of radius 0.8 arcsec. 
These results show that the effect of SN contamination on the \imfit\ fits is quite small for most objects (Fig. \ref{fig:llSNmask}). \linclacc\  values agree to within $\pm 0.1$ ($\pm 0.05$) for 90 (80) per cent of the primary sample; none of the conclusions of the paper are affected. Most of the deviant points are centrally located supernovae, as is also the case in a comparison of masked and unmasked $B/T$ values. 

\subsection{SN Positional Errors}
\label{subsec:xyerr}

As mentioned in \S \ref{subsec:hostfitting}, the centroids of SNe and their hosts are not strongly affected by seeing. For single SDSS scans, the centroiding error is $\sigma_{xy}\simeq 0.1$ arcsec, both for stars at $r=22$ mag (the brightness of our faintest SNe), and also for galaxies at $r=20$ mag (our faintest hosts) \citep{Pier2003}. These errors will become smaller in the stacked images.

We have simulated the effects of this positional error on \linclacc\  for both disk and bulge components. The size of the effect depends on $\sigma_{xy}/r_{eff}$ and on S\'ersic index $n$. For small disk galaxies ($r_{eff}=0.8$ arcsec), the effect is negligible; for bulges with the same $r_{eff}$, there is a deficit of objects at $L_{inc}/L_{acc} \la 0.1$, which are scattered outwards from the cores of bulges. We therefore try restricting the bulge analysis to $L_{inc}/L_{acc} \ge 0.1$ in \S\ref{sec:bulges}. 

\subsection{Bulge Structure}
\label{subsec:bulgen}

We assumed $n=4$ for the bulge component of the 2 component fits -- i.e. that bulges are ``mini-ellipticals'' that follow the de Vaucouleurs profile (e.g. \citealt{Kormendy2004}). In fact, there exists strong evidence that bulges and pseudo-bulges possess a wide range of S\'ersic $n$ values, ranging from disk-like $n\sim 1$ (pseudo-bulges), to $2 \la n \la 6$ for classical bulges (\citealt{Kormendy2004,Andredakis1995,Fisher2008,Gadotti2009,Gao2020}). The bulge of the Milky Way (Sbc) is close to $n=1$ (e.g. \citealt{Kent1991}); the bulge of the nearby Sb spiral M31 possesses $n\simeq 2.2$ \citep{Courteau2011}.

From the above literature several things are clear. Galaxies with prominent bulges ($B/T \ga 0.2$, roughly Sb or earlier) tend to have larger $n_{bulge}$ (though note the different $n$ distributions for classical bulges in \citealt{Fisher2008} and \citealt{Gao2020}), and there is a clear trend of increasing $n_{bulge}$ with $B/T$. Galaxies with $B/T > 0.2$ are the systems discussed in \S\ref{sec:bulges}. Luminous galaxies (such as in magnitude limited surveys of SN hosts -- e.g. \S \ref{subsec:sample}) also tend to have large $n_{bulge}$. Given that we cannot fit $n_{bulge}$ from our data, $n_{bulge}=4$ is a reasonable starting point. Our approach in \S\ref{sec:bulges} will be to see which value of $n_{bulge}$ best fits deviations from a straight line in the \linclacc\  plot calculated with $n_{bulge}=4$. 

What is the effect of using an incorrect value of $n_{bulge}$ on our estimated value of $B/T$? Surprisingly, not much. We have used \imfit\ to fit seeing degraded models of M31 ($n_{bulge}^{true}=2.2, n_{bulge}^{fit}=4$) redshifted to $z=0.1-0.3$, and find that $B/T$ is recovered to $\pm0.03$. This is because, at these redshifts, bulges are at best poorly resolved. Errors become larger for $z<0.1$, seeing FWHM $<0.5$ arcsec, or for large bulges.

Finally we note that, because of resolution effects, nuclei and bulges are difficult to distinguish in our data. Galaxies with bright nuclei will be fitted with an overly bright bulge and an $r_{eff}$ value that is too small.

\vfill\eject

\bibliography{SDSSSNE_AJv2,CP}
\bibliographystyle{aasjournal}

\end{document}